# Mapping mesoscopic phase evolution during e-beam induced transformations via deep learning of atomically resolved images

R. K. Vasudevan[1,2,*], N. Laanait[3], E. M. Ferragut[4,‡], K. Wang[1,2], D. Geohegan[1,2], K. Xiao[1,2], M. Ziatdinov[1,2], S. Jesse[1,2], O. Dyck[1,2] and S. V. Kalinin[1,2]

[1]Center for Nanophase Materials Sciences, [2]Institute for Functional Imaging of Materials and [3]Computational Sciences and Engineering Division and [4]Quantum Computing Institute,
Oak Ridge National Laboratory, Oak Ridge TN, 37831, USA
[‡]Current affiliation: UnitedHealth Group

**Abstract**

Understanding transformations under electron beam irradiation requires mapping the structural phases and their evolution in real time. To date, this has mostly been a manual endeavor comprising of difficult frame-by-frame analysis that is simultaneously tedious and prone to error. Here, we turn towards the use of deep convolutional neural networks (DCNN) to automatically determine the Bravais lattice symmetry present in atomically-resolved images. A DCNN is trained to identify the Bravais lattice class given a 2D fast Fourier transform of the input image. Monte-Carlo dropout is used for determining the prediction probability, and results are shown for both simulated and real atomically-resolved images from scanning tunneling microscopy and scanning transmission electron microscopy. A reduced representation of the final layer output allows to visualize the separation of classes in the DCNN and agrees with physical intuition. We then apply the trained network to electron beam-induced transformations in $WS_2$, which allows tracking and determination of growth rate of voids. These results are novel in two ways: (1) It shows that DCNNs can be trained to recognize diffraction patterns, which is markedly different from the typical 'real image' cases, and (2) it provides a method with in-built uncertainty quantification, allowing the real-time analysis of phases present in atomically resolved images.

**E-mail:** *vasudevanrk@ornl.gov

**Introduction**

Phase formation and evolution is one of the key processes in the solid-state chemistry, physics, electrochemistry, and materials science. Correspondingly, studies of the kinetics and thermodynamics of phase evolution play a special role in virtually all areas of modern science, ranging from battery and fuel cell operation to formation and operation of electro-resistive devices to solid state reactions involved in materials formation.[1-5] Despite the cornerstone importance of this field, information on process parameters is derived from macroscopic measurements, or via scattering studies of large volume of material, while atomistic mechanisms remain largely unexplored.

In the last several years, advances in (scanning) transmission electron microscopy ((S)TEM) have enabled visualization the atomic dynamics during multiple solid state processes,[6,7] including vacancy ordering in cobaltites,[8] phase transformation at crystalline oxide interfaces,[9] restructuring in 2D silica glass,[10] single point defect motion,[11-13] single atom catalytic activity,[14-16] elastic-plastic transitions,[17] and dislocation migration.[18] The atomic dynamics were observed in response to classical macroscopic stimuli such as temperature and pressure, but also as a response to e-beam irradiation.[19,20] Understanding these transformations requires establishing the nature of the phases and structures contained in a single image, ideally during image acquisition process. These studies are particularly important in the context of the e-beam manipulation of matter atom by atom, as recently demonstrated.[21-26]

For any atomically-resolved image, the first task is to identify the constituent lattice types present and determine their spatial distribution in the image. A review of existing approaches by Moeck is a good reference on this topic.[27] These codes and software tools analyze individual images and (some) require the user to select the constituent unit cell, and then proceed to use motif-matching style algorithms to determine the space group. As outlined in the review[27], there are several issues that limit the application of these methods, most obviously the difficulty in determining the uncertainties in the classification associated with the noise in atomic positions, but also that in many images a mixture of phases is present, necessitating an image segmentation that respects the available symmetries.

Previously we have used a sliding window-based method, to segment images based on linear unmixing of the (local) 2D fast Fourier Transform (FFT) spectra, via the N-FINDR unmixing algorithm[28] with both positivity and sum-to-one constraints.[29,30] Ideally one would like to impose an additional constraint that the endmembers should conform to some symmetry types in the case of atomically-resolved images. For the 2D case, each lattice must reside in one of the five Bravais Lattice types, i.e. square, rectangle, centered rectangle, hexagonal and rhombohedral. Thus, this task effectively boils down to determination of which of the five Bravais lattice types each portion of the image belongs to. This can be trivially defined mathematically for the standard coordinate frame, but the task is more complex when one considers that in any real atomically resolved image, the lattices themselves may be rotated in arbitrary fashion, and multiple diffraction orders can be present. Moreover, there will always exist noise and slight distortions that can make hand-crafting the particular rules (i.e., the features) more difficult in practice.

In the past, similar problems were faced by the computer vision community, where approaches such as the scale invariant feature transform[31] were designed as methods to attempt to determine the appropriate feature vectors that would be suitable for the classification. However, such methods have quickly become overtaken by the development of deep convolutional neural networks (DCNN), which have displayed remarkable accuracy that rivals humans in image classification tasks.[32,33] The key advantage of a DCNN is the ability for the convolutional layers to 'learn' abstract features that are mostly independent of position and scale, allowing them to be useful in predictions.[34] Interestingly, whether a DCNN can be used in the determination of crystal structure is unknown, because in this case, the features (diffraction spots) are identical, but the classification hinges on their relative positions and angles. In theory, this is a poorly suited task for fully convolutional neural networks, given that each layer only received input from the node immediately previous.

Here, we show that through the appropriate training and network architecture, a DCNN can learn the five different crystal structure types, independent of the scale and rotation, and allow for automated prediction of the Bravais lattice type. We find that the network has an 85% success rate on the simulated validation set, and that in 75% of cases where the prediction is incorrect, the second-most likely prediction is correct. We further validate the DCNN on real images obtained from both scanning tunneling microscopy and scanning transmission electron microscopy, and utilize the Monte-Carlo dropout[35] technique to estimate the confidence of the predictions. We visualize the network with a reduced t-SNE representation, which surprisingly maps the symmetry classes in a physically intuitive form. We then apply our model towards understanding the phase transformation in Mo doped $WS_2$ under electron beam irradiation, which allows us to determine the phase transformation kinetics, which are in this case appear exponential with beam time exposure. Combined with other recent advances in deep learning for atomic scale image analysis[36], we believe these studies lay the foundation for a 2D AI crystallographer, which will be integral in future automated analysis workflows for STEM and STM platforms.

**Training Set**

The key task of machine learning based approaches for physics-related studies is to establish appropriate training set that captures the physics of the problem. For the image set for our DCNN, we generated a set of images of six different lattice types, with 4000 members in each set totaling 24000 images. Note that the sixth set was a 'noise' class, i.e. what the classifier is expected to return if there is no periodicity present (or no atoms are present). For each Bravais lattice type we vary all possible lattice parameters (effectively allowing the DCNN to learn that the lattice types are invariant on scale), and then add arbitrary rotations, along with some randomization of atomic positions, to simulate (uncorrelated) disorder. Finally, the FFT of each image (only amplitude) is taken and stored. Note that for the noise class, we generated images by beginning with a rhombohedral lattice, and then perturbing the atomic positions until no clear FFT diffraction spots could be observed. In contrast to most work in the deep learning community, we worked exclusively in the Fourier space, because this type of preprocessing is expected to maximize information on the periodicity, and suppress noise to the central reflection, simplifying the classification task. At the same time, this allows more transferability of the approach, as the FFT is expected to be similar across different instrumentation platforms, negating the need for specific simulations (such as multi-slice for STEM, or density functional theory for STM).

**Network Architecture**

The architecture of the DCNN is shown in Fig. 1(a), and consists of 3 convolutional layers, followed by an average pooling layer, and then a dense (fully connected) layer. An average pooling layer was added to increase the 'connectedness' of the convolutional filters, matching the physics of the problem. Rectified linear units were used as activation functions in these layers. Finally, the last layer consists of a six-unit dense layer with softmax activation, so that the classification outputs would sum to one, respecting the choices available. It is this layer that provides the estimate of which class the input image belongs to. To prevent overfitting, we employed the commonly used weight decay ($l_2$ regularization), as well as dropout layers (shown in red), which were set to randomly mask 20% of the input from the previous layer, before propagating them to the next layer on each training batch. In addition to boosting the generalizability of the network, dropout also allows for uncertainty quantification, as shown by Gal and Ghahramani.[35] Effectively, running a single image through the *trained* network thousands of times with the dropout layers active can serve as a type of inference, allowing probability distributions of the resulting classifications to be computed. Optimization of the network was performed with the Adam optimizer utilizing the cross-entropy metric. All work was conducted using keras with a TensorFlow backend. We trained on 19200 images and validated on 4,800 for a total of 30 epochs (one epoch is one complete pass through the training dataset), with the results of the training accuracy and validation accuracy shown in Fig. 1(b). It is seen that after ~15 epochs, little improvement in the validation accuracy occurs, which plateaus at about 85%.

**Results**

We first plot some results of the network from the validation set, i.e., the data that the network has not seen during the training step. The results are shown in Figure 2, which shows that the network misclassified 5 of the 18 images presented, highlighted in red (more examples shown in Supplementary Figure 1). These validation images show that the DCNN is indeed learning that scale is not an important feature for the classification task, as can be seen by the FFT spectra of the centered rectangle lattices of different scale. Unlike in a standard image classification task, this task is more difficult given that some classes can merge into others, for example a rectangle can easily morph into the square class for values of lattice parameters that are (within noise) indistinguishable; therefore, 100% accuracy is neither possible nor desirable. Instead, we computed the probability over the classifications via the use of the Monte-Carlo dropout method[35], with 5000 iterations for each image, with the probability shown next to each prediction in the titles in Figure 2. This allows us to determine a mean probability and a standard deviation over the predictions. Given the difficulty of the task, it is more important that the confidence estimates be appropriate, and that the second-most likely physical solution be correct. It is seen that the confidence of the incorrect predictions is generally low (below 65% in all cases) and the misclassifications generally appear reasonable.

**Errors and Noise Limits**

To investigate these errors more closely, we chose 1000 images from the training set, and investigated the statistics during incorrect predictions. The prediction confidence histogram is plotted in Fig. 3(a) and shows a distribution that is centered around lower values, i.e. when the prediction is wrong, the confidence tends to be low. This is of course evidence that the network is uncertain, as should be the case (one would not wish a network to be confident in incorrect predictions). For more inspection, we turned to an example of an incorrect prediction for a single lattice, shown in Fig. 3(b). The prediction confidence for the six classes along with their standard deviations are plotted in Fig. 3(c) for this image. As can be seen, the probability is highest for the rectangle class, but the second-most likely prediction is square, which is the ground truth. The suggests a good degree of robustness for the classifier (more examples shown in Supplementary Figure 2).

Of importance is the DCNN's susceptibility to noise. As an example, we plot a (noisy) STM image of graphene in Fig. 3(d). We proceeded to take the 2D FFT of this image using 3 different window sizes: the first window was the size of the entire image, the next two were the sizes of the two red squares shown in Fig. 3(c). That is, the window size was decreased by a factor of 2 each time, and the DCNN prediction on the symmetry class was made. The results are seen in Fig. 3(e) and show that a rhombohedral symmetry class is predicted for the largest FFT (shown inset in Fig. 3(e)), but that the confidence in this prediction steadily decreases with decreasing window sizes. The associated FFT spectra become much more convolved, and substantial edge effects can be seen. It is interesting that the rhombohedral class, instead of the hexagonal class, is predicted for this image. The lattice is ideally hexagonal, but substantial drift and disorder effects serve to reduce the symmetry to the rhombohedral class. Moreover, the decreasing confidence of the classifier with smaller window sizes is instructive of the degree to which features should be distinguishable for the DCNN to accurately make a symmetry judgement.

**Visualizing the features**

To gain more insight into how the DCNN is working, we may observe that the purpose of a DCNN is to learn features of image data that can then be separated into the available classes via a linear classifier. Thus, observing the weights of the last layer (before the classification layer) can provide some information as to the nature of this classification space. Given that this is a 128-length vector, visualization requires projecting to a lower dimensional space. However, this dimensionality reduction should be done via a method that maintains (approximately) the distance between images in both the low and high-dimensional representations. The t-distributed stochastic neighbor embedding (t-SNE)[37] provides such a method, and we used it to calculate the 2D embedding for 1000 training images. These vectors then form the ($x$,$y$) coordinates for each image, as shown in Fig 4. The specific images are colored by their true classes. Essentially, images that are nearby in this representation are also nearby in the DCNN and can be though to be 'similar' as far as the network is concerned. Note that the noise class is substantially further away than has been drawn in the figure. Interestingly, the network separated the centered rectangle class very distinctly from the hexagonal class. Moreover, the rhombohedral class members can be seen scattered near the other classes, as would be expected (given this is the lowest symmetry class).

The closeness of the square and rectangle classes is also expected, and altogether support the idea that the classification appears to be made by the DCNN based on symmetry features.

**Application to Edge Cases**

To better determine how the classifier separates edge cases (i.e., cases where the symmetry is close to other classes), we tested the DCNN on two images of graphene taken by scanning tunneling microscopy. In the first image, show in Fig. 5(a), a slight drift of the microscope is evident. Thus, this provides an excellent test case for the DCNN to determine the stringency of the constraints for the network to classify a structure as hexagonal. The 2D FFT of the same STM image is shown in Fig. 5(b), and the probabilities of classification are shown in Fig. 5(c) (note, 5000 passes of the network were used to determine the probabilities, which takes ~1 minute per image on a standard desktop). The standard deviation is shown as error bars on the predictions. The DCNN predicts, that the lattice type is rhombohedral. It then predicts that the second most likely structure is the hexagonal class, and there are relatively large uncertainties in these predictions. The reason for the rhombohedral classification as opposed to hexagonal may be due to the slight disorder (reflected as dispersion of the diffraction spots). To test this assumption, we used the DCNN to classify the image in Fig. 5(d,e), again of graphene, but this time containing defects as well a larger scale Moiré pattern. The FFT spectra is shown in Fig. 3(e) and shows the higher order reflections. Interestingly, the classifier is robust enough to determine the hexagonal symmetry in this instance, despite the extra reflections. Indeed, hand-coding features is highly likely to fail for cases such as these, due to multiple spots having similar intensity but being of different orders.

**Extension to Transformations**

While interesting for single images, the real utility of these classifiers is undoubtedly for image sequences (movies), comprising the evolution of atomic structures under the effect of temperature or electron beam bombardment, where such analysis cannot be done by simple observation. We captured a movie of e-beam-mediated decomposition of Mo-doped $WS_2$, with frames 30,60 and 90 shown in Fig. 6(a-c), respectively. Here, we used a 100 kV beam which will produce damage from both radiolysis and knock-on damage particularly to the S atoms.

We calculated the maximum energy transfer via the formula

$$\Delta E = \frac{2(E + 2E_0)E}{Mc^2} \qquad (1)$$

where $E$ is the beam electron kinetic energy, $E_0 = m_0 c^2$ (511 keV), $m_0$ is the rest mass of the electron, and M is the atom mass, which yields 1.6 eV for W and 7.5 eV for S.[38] Based on these estimates and average binding energy of atoms in the lattice, the e-beam effect can be represented as a gradual reduction of material and oversaturation with respect to S vacancies, leading to the gradual collapse of the 2D lattice and formation of W-rich phases. Moreover, previous studies have found evidence for e-beam induced chemical reactions,[39,40] which may also lead to a damage-assisting mechanism here as well. Thus, unravelling the atomic level details of the damage evolution is a non-trivial task, yet understanding and controlling such evolution on an atom-by-atom basis may enable atomic scale defect generation and manipulation akin to that being developed for graphene.[6,7,12,21,22,25,41]

We therefore investigate an automated routine for identifying phase evolution at the atomic scale. For each frame, we computed the local FFT spectra via the sliding window method,[30] with a window size of 128px and step size of 16px (the outline of the window is shown in Fig. 6(a) in white). Running the DCNN on each window allows classification of the local symmetries present in each frame of the image sequence, and is plotted in Fig 6(d,e) for the 90$^{th}$ frame. Full results for the movie are shown in Supplementary Video 1. The mean % of classification of each Bravais lattice symmetry class is shown in Fig 6(f) for each frame. It should be noted that this is simply the mean of the probabilities of individual windows of each frame belonging to a particular class and does not necessarily imply the existence of voids. For instance, if the probability of the void is 10% in each of the windows of the frame, but the probability of rhombohedral lattice is 90% in each window, then the mean probability of the noise class is still 10%, even though the DCNN has determined that all the windows belong to the rhombohedral class. Nonetheless, we expect this to be valid here, as the probability of the noise class would increase if many windows began to be classified as such within the frame. Due to the large scale random motions of the atoms, the symmetry defaults to rhombohedral. Initially there is a steady-state situation wherein the beam is not introducing defects (voids) large enough to cause cascade and runaway growth of the voids. Eventually, voids on the bottom right of the frame become larger, leading to runaway growth (see progression of frames in Fig. 6(a-c).

From Fig. 6(c), we observe the exponential increase in the 'noise' phase fraction in Fig. 6(f) beyond a particular time. We can express this mathematically as $\frac{\partial A(t)}{\partial t} = \rho A(t)$, where $A(t)$ is the area of the hole which changes with time, and $\rho$ is the sputtering rate for the atoms at the hole edge. Here, we are simply saying that the hole growth rate is dependent on the size of the holes. This is an ordinary differential equation with the solution, $A(t) = A_0 \exp(\rho t)$ where $A_0$ is the initial hole size. Note here the sizes $A(t)$ are given as phase fractions, i.e. between 0 and 1. Clearly, the initial hole size cannot be zero, which suggests that this model is only useful for describing what happens *after* a small hole has been formed. Nevertheless, fitting this model to the data we can obtain an estimate for the initial hole size and determine the sputtering rate during our experiment. However, because this equation describes only the growth of the holes and not the generation, we must add another constant which represents the initial state where holes are being generated and healed randomly under the electron beam. In this situation, holes are indeed present, so our hole fraction does not start at zero, i.e. $A(t) = A_0 \exp(\rho t) + C$. We use the first 100s as a representation of this initial state to determine the value of this constant, $C$, which is ~0.059. We then fit the above equation to the observed data, allowing $A_0$ and $\rho$ to be our fitting parameters for time t>100s. The curve of best fit is represented in Fig. 6(f) as a dashed black line. We see that our model closely follows the observed data and we obtain a minimum hole size of $1.85 \times 10^{-5}$ (equivalent to an area of ~0.004nm$^2$), which is approximately the dimensions of a single vacancy, and a sputtering rate of ~9 nm$^2$/sec. As a sanity check, we also expect to observe agreement in this case between the DNNC and a simple threshold on the image (since the holes are clearly very dark). Discriminating the hole fraction through intensity thresholding results in the black dots in Fig. 6(f). While the fractions are slightly different, they are generally within 20% of the DCNN, and further, there is good agreement as the voids become larger. These results suggest that the DNNC method is robust and reliable.

**Discussion**

Here, we introduce the deep learning model for the identification of symmetry classes in atomically resolved images in electron and scanning probe microscopies. This task is highly non-trivial given that local symmetry is generally poorly defined given the noise in the images and must allow for image rotations and scale invariance. The DCNN overcomes these problems and provides a robust tool for analysis of images and dynamic image sequence data. The future development of this field can include combinations of the DCNN with local atomic finding that will allow determination of primitive cells. We believe these approaches can be complementary, with symmetry finders improving the atomic recognition by providing priors based on expected local symmetries, and in turn improving symmetry determination. This approach and provide way to fully unravel the mechanisms involved in atomic changes during temperature- and beam induced processes.

Further, this data can be of significant interest in the context of recent advances in understanding and designing new materials via large scale integration of predictive modelling and machine learning methods,[42-49] collectively referred to as the Materials Genome.[46,50,51] This progress hinges on the fact that crystalline systems are associated with long-range periodicity and symmetries, which form the natural descriptors of crystalline materials. Combination of the structural data, processing parameters, and functionalities further enabled artificial intelligence driven workflows for materials design and even optimization of synthesis,[52-55] extending recent advances for organic molecule synthesis[56,57] towards inorganic systems. We believe that approaches developed here and in other publications[36,58,59] will provide experiment-based descriptors for structure of solids that can be used in similar workflows.

**Acknowledgements**

The work was supported by the U.S. Department of Energy, Office of Science, Materials Sciences and Engineering Division (R.K.V., S.V.K., M.Z.). The synthesis of 2D materials was supported by the U.S. Department of Energy, Office of Science, Materials Sciences and Engineering Division (K.W., K.X., D.B.G.). This research was conducted and partially supported (S.J.) at the Center for Nanophase Materials Sciences, which is a US DOE Office of Science User Facility. N.L. acknowledges support from the Eugene P. Wigner Fellowship program at Oak Ridge National Lab.

# References

1. Gu, M. *et al.* Formation of the spinel phase in the layered composite cathode used in Li-ion batteries. *ACS nano* **7**, 760-767 (2012).
2. Kilner, J. A. & Burriel, M. Materials for intermediate-temperature solid-oxide fuel cells. *Annual Review of Materials Research* **44**, 365-393 (2014).
3. Finegan, D. P. *et al.* In-operando high-speed tomography of lithium-ion batteries during thermal runaway. *Nat. Commun.* **6** (2015).
4. Sun, Y. *et al.* In-operando optical imaging of temporal and spatial distribution of polysulfides in lithium-sulfur batteries. *Nano Energy* **11**, 579-586 (2015).
5. Sebastian, A., Le Gallo, M. & Krebs, D. Crystal growth within a phase change memory cell. *Nat. Commun.* **5**, 4314 (2014).
6. Mishra, R., Ishikawa, R., Lupini, A. R. & Pennycook, S. J. Single-atom dynamics in scanning transmission electron microscopy. *MRS Bull.* **42**, 644-652, doi:10.1557/mrs.2017.187 (2017).
7. Zhao, X. *et al.* Engineering and modifying two-dimensional materials by electron beams. *MRS Bull.* **42**, 667-676, doi:10.1557/mrs.2017.184 (2017).
8. Kim, Y. M. *et al.* Probing oxygen vacancy concentration and homogeneity in solid-oxide fuel-cell cathode materials on the subunit-cell level. *Nat. Mater.* **11**, 888-894 (2012).
9. Jesse, S. *et al.* Atomic-Level Sculpting of Crystalline Oxides: Toward Bulk Nanofabrication with Single Atomic Plane Precision. *Small* **11**, 5895-5900, doi:10.1002/smll.201502048 (2015).
10. Huang, P. Y. *et al.* Imaging Atomic Rearrangements in Two-Dimensional Silica Glass: Watching Silica's Dance. *Science* **342**, 224-227, doi:10.1126/science.1242248 (2013).
11. Kotakoski, J., Mangler, C. & Meyer, J. C. Imaging atomic-level random walk of a point defect in graphene. *Nature Communications* **5**, 3991, doi:10.1038/ncomms4991 https://www.nature.com/articles/ncomms4991#supplementary-information (2014).
12. Susi, T. *et al.* Silicon-Carbon Bond Inversions Driven by 60-keV Electrons in Graphene. *Phys. Rev. Lett.* **113**, 115501 (2014).
13. Ishikawa, R. *et al.* Direct Observation of Dopant Atom Diffusion in a Bulk Semiconductor Crystal Enhanced by a Large Size Mismatch. *Phys. Rev. Lett.* **113**, 155501 (2014).
14. Wang, W. L. *et al.* Direct Observation of a Long-Lived Single-Atom Catalyst Chiseling Atomic Structures in Graphene. *Nano Lett.* **14**, 450-455, doi:10.1021/nl403327u (2014).
15. Zhao, J. *et al.* Direct in situ observations of single Fe atom catalytic processes and anomalous diffusion at graphene edges. *Proceedings of the National Academy of Sciences* **111**, 15641-15646, doi:10.1073/pnas.1412962111 (2014).
16. Ta, H. Q. *et al.* Single Cr atom catalytic growth of graphene. *Nano Research*, doi:10.1007/s12274-017-1861-3 (2017).
17. Dai, S. *et al.* Electron-Beam-Induced Elastic–Plastic Transition in Si Nanowires. *Nano Lett.* **12**, 2379-2385, doi:10.1021/nl3003528 (2012).
18. Robertson, A. W. *et al.* Partial Dislocations in Graphene and Their Atomic Level Migration Dynamics. *Nano Lett.* **15**, 5950-5955, doi:10.1021/acs.nanolett.5b02080 (2015).

19      Lin, Y.-C., Dumcenco, D. O., Huang, Y.-S. & Suenaga, K. Atomic mechanism of the semiconducting-to-metallic phase transition in single-layered MoS 2. *Nat. Nanotech.* **9**, 391 (2014).
20      Jiang, N. Electron beam damage in oxides: a review. *Rep Prog Phys* **79**, 016501 (2015).
21      Susi, T., Meyer, J. C. & Kotakoski, J. Manipulating low-dimensional materials down to the level of single atoms with electron irradiation. *Ultramicroscopy* **180**, 163-172 (2017).
22      Dyck, O., Kim, S., Kalinin, S. V. & Jesse, S. Placing single atoms in graphene with a scanning transmission electron microscope. *Appl. Phys. Lett.* **111**, 113104, doi:10.1063/1.4998599 (2017).
23      Jesse, S. *et al.* Direct atomic fabrication and dopant positioning in Si using electron beams with active real time image-based feedback. *arXiv preprint arXiv:1711.05810* (2017).
24      Jesse, S. *et al.* Patterning: Atomic-Level Sculpting of Crystalline Oxides: Toward Bulk Nanofabrication with Single Atomic Plane Precision (Small 44/2015). *Small* **11**, 5854-5854 (2015).
25      Susi, T. *et al.* Towards atomically precise manipulation of 2D nanostructures in the electron microscope. *2D Materials* **4**, 042004 (2017).
26      Dyck, O., Kim, S., Kalinin, S. V. & Jesse, S. E-beam manipulation of Si atoms on graphene edges with aberration-corrected STEM. *arXiv preprint arXiv:1710.10338* (2017).
27      Moeck, P. Towards generalized noise-level dependent crystallographic symmetry classifications of more or less periodic crystal patterns. *arXiv preprint arXiv:1801.01202* (2018).
28      Winter, M. E. in *SPIE's International Symposium on Optical Science, Engineering, and Instrumentation.*  266-275 (International Society for Optics and Photonics).
29      Vasudevan, R. K., Ziatdinov, M., Jesse, S. & Kalinin, S. V. Phases and interfaces from real space atomically resolved data: Physics-based deep data image analysis. *Nano Lett.* **16**, 5574-5581 (2016).
30      Vasudevan, R. K. *et al.* Big data in reciprocal space: Sliding fast Fourier transforms for determining periodicity. *Appl. Phys. Lett.* **106**, 091601 (2015).
31      Lowe, D. G. in *Computer vision, 1999. The proceedings of the seventh IEEE international conference on.*  1150-1157 (Ieee).
32      LeCun, Y., Bengio, Y. & Hinton, G. Deep learning. *Nature* **521**, 436-444 (2015).
33      Simonyan, K. & Zisserman, A. Very deep convolutional networks for large-scale image recognition. *arXiv preprint arXiv:1409.1556* (2014).
34      LeCun, Y. & Bengio, Y. Convolutional networks for images, speech, and time series. *The handbook of brain theory and neural networks* **3361**, 1995 (1995).
35      Gal, Y. & Ghahramani, Z. in *International conference on machine learning.*  1050-1059.
36      Ziatdinov, M. *et al.* Deep Learning of Atomically Resolved Scanning Transmission Electron Microscopy Images: Chemical Identification and Tracking Local Transformations. *ACS Nano*, doi:10.1021/acsnano.7b07504 (2017).
37      Maaten, L. v. d. & Hinton, G. Visualizing data using t-SNE. *Journal of Machine Learning Research* **9**, 2579-2605 (2008).
38      Nan, J. Electron beam damage in oxides: a review. *Rep. Prog. Phys.* **79**, 016501 (2016).

39  Wei, X., Wang, M.-S., Bando, Y. & Golberg, D. Electron-Beam-Induced Substitutional Carbon Doping of Boron Nitride Nanosheets, Nanoribbons, and Nanotubes. *ACS Nano* **5**, 2916-2922, doi:10.1021/nn103548r (2011).
40  Ramasse, Q. M. *et al.* Direct Experimental Evidence of Metal-Mediated Etching of Suspended Graphene. *ACS Nano* **6**, 4063-4071, doi:10.1021/nn300452y (2012).
41  Kalinin, S. V. & Pennycook, S. J. Single-atom fabrication with electron and ion beams: From surfaces and two-dimensional materials toward three-dimensional atom-by-atom assembly. *MRS Bull.* **42**, 637-643, doi:10.1557/mrs.2017.186 (2017).
42  Ramakrishnan, R., Dral, P. O., Rupp, M. & von Lilienfeld, O. A. Quantum chemistry structures and properties of 134 kilo molecules. *Scientific data* **1**, 140022, doi:10.1038/sdata.2014.22 (2014).
43  Bartok, A. P., Gillan, M. J., Manby, F. R. & Csanyi, G. Machine-learning approach for one- and two-body corrections to density functional theory: Applications to molecular and condensed water. *Physical Review B* **88**, doi:054104 10.1103/PhysRevB.88.054104 (2013).
44  Rupp, M., Tkatchenko, A., Muller, K. R. & von Lilienfeld, O. A. Fast and Accurate Modeling of Molecular Atomization Energies with Machine Learning. *Physical Review Letters* **108**, doi:058301 10.1103/PhysRevLett.108.058301 (2012).
45  Raghunathan Ramakrishnan, P. O. D., Matthias Rupp, O. Anatole von Lilienfeld. Big Data Meets Quantum Chemistry Approximations: The Δ-Machine Learning Approach. *J. Chem Theory Comput* **11**, 2087-2096 (2015).
46  Curtarolo, S. *et al.* The high-throughput highway to computational materials design. *Nat. Mater.* **12**, 191-201, doi:10.1038/nmat3568 (2013).
47  Setyawan, W. & Curtarolo, S. High-throughput electronic band structure calculations: Challenges and tools. *Comput. Mater. Sci.* **49**, 299-312, doi:10.1016/j.commatsci.2010.05.010 (2010).
48  Jain, A. *et al.* A high-throughput infrastructure for density functional theory calculations. *Comput. Mater. Sci.* **50**, 2295-2310, doi:10.1016/j.commatsci.2011.02.023 (2011).
49  Fischer, C. C., Tibbetts, K. J., Morgan, D. & Ceder, G. Predicting crystal structure by merging data mining with quantum mechanics. *Nat. Mater.* **5**, 641-646, doi:10.1038/nmat1691 (2006).
50  <https://darkreactions.haverford.edu/> (
51  Jain, A. *et al.* Commentary: The Materials Project: A materials genome approach to accelerating materials innovation. *APL Mater.* **1**, doi:011002 10.1063/1.4812323 (2013).
52  Michel, K. & Meredig, B. Beyond bulk single crystals: A data format for all materials structure-property-processing relationships. *Mrs Bulletin* **41**, 617-622, doi:10.1557/mrs.2016.166 (2016).
53  O'Mara, J., Meredig, B. & Michel, K. Materials Data Infrastructure: A Case Study of the Citrination Platform to Examine Data Import, Storage, and Access. *Jom* **68**, 2031-2034, doi:10.1007/s11837-016-1984-0 (2016).
54  Sparks, T. D., Gaultois, M. W., Oliynyk, A., Brgoch, J. & Meredig, B. Data mining our way to the next generation of thermoelectrics. *Scr. Mater.* **111**, 10-15, doi:10.1016/j.scriptamat.2015.04.026 (2016).

55  Hill, J. *et al.* Materials science with large-scale data and informatics: Unlocking new opportunities. *Mrs Bulletin* **41**, 399-409, doi:10.1557/mrs.2016.93 (2016).
56  Cadeddu, A., Wylie, E. K., Jurczak, J., Wampler-Doty, M. & Grzybowski, B. A. Organic Chemistry as a Language and the Implications of Chemical Linguistics for Structural and Retrosynthetic Analyses. *Angew. Chem.-Int. Edit.* **53**, 8108-8112, doi:10.1002/anie.201403708 (2014).
57  Szymkuc, S. *et al.* Computer-Assisted Synthetic Planning: The End of the Beginning. *Angew. Chem.-Int. Edit.* **55**, 5904-5937, doi:10.1002/anie.201506101 (2016).
58  Ziatdinov, M. *et al.* Deep analytics of atomically-resolved images: manifest and latent features. *arXiv preprint arXiv:1801.05133* (2018).
59  Liu, P. *et al.* A deep learning approach to identify local structures in atomic-resolution transmission electron microscopy images. *arXiv preprint arXiv:1802.03008* (2018).

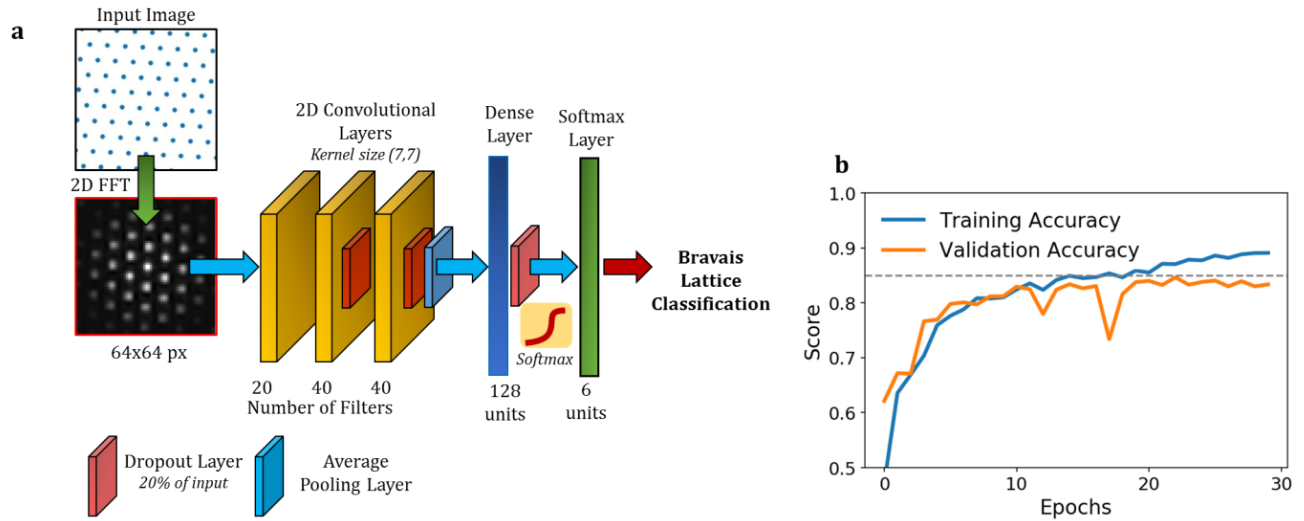

**Figure 1: Deep Convolutional Neural Network for symmetry classification.** (**a**) Schematic of the DCNN structure. A lattice image is input, and transformed via a 2D fast Fourier Transform (FFT). This image is input to the DCNN, which outputs probability of classification into one of the six classes (five Bravais lattice types, and one for 'noise', i.e., no periodicity). (**b**) Training and validation accuracy as a function of epoch. One epoch is one complete pass through the training data.

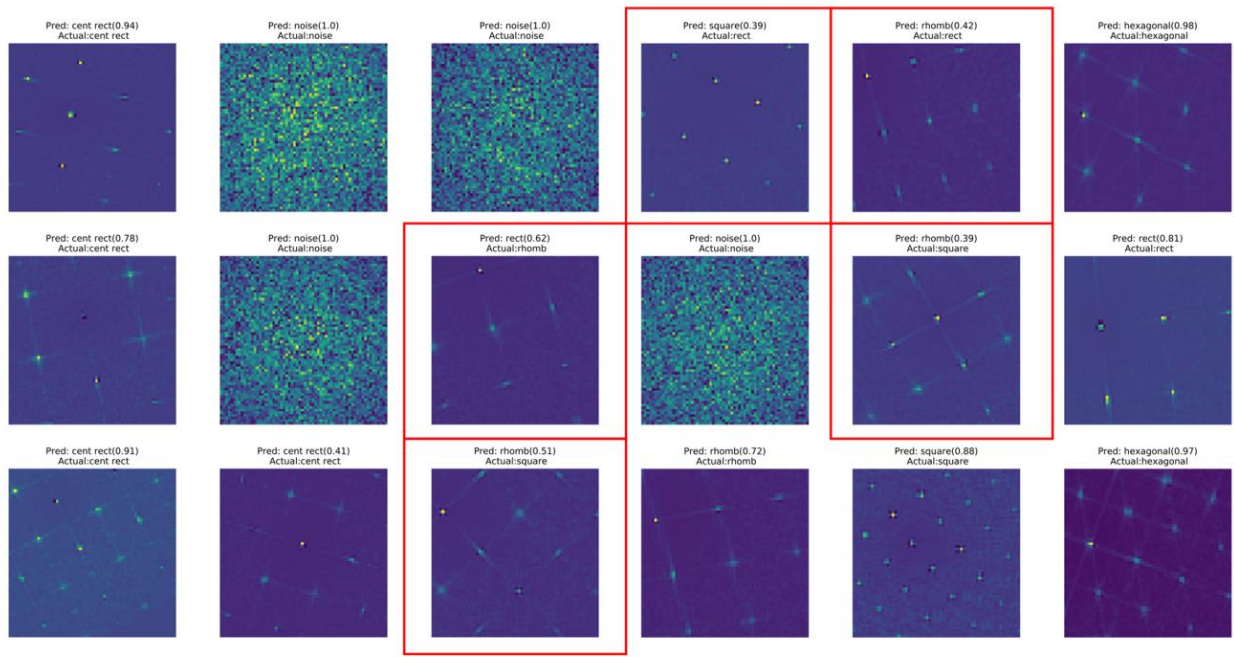

**Figure 2: Classification examples of the DCNN.** Randomly selected (simulated) images from the validation set, and their classification by the DCNN with the predicted class ('Pred:') shown above the true ('Actual') class. The probability is shown in parentheses next to the prediction. Note the probabilities are computed via 5000 passes of the individual images through the network with dropout layers active. Misclassified images are boxed in red.

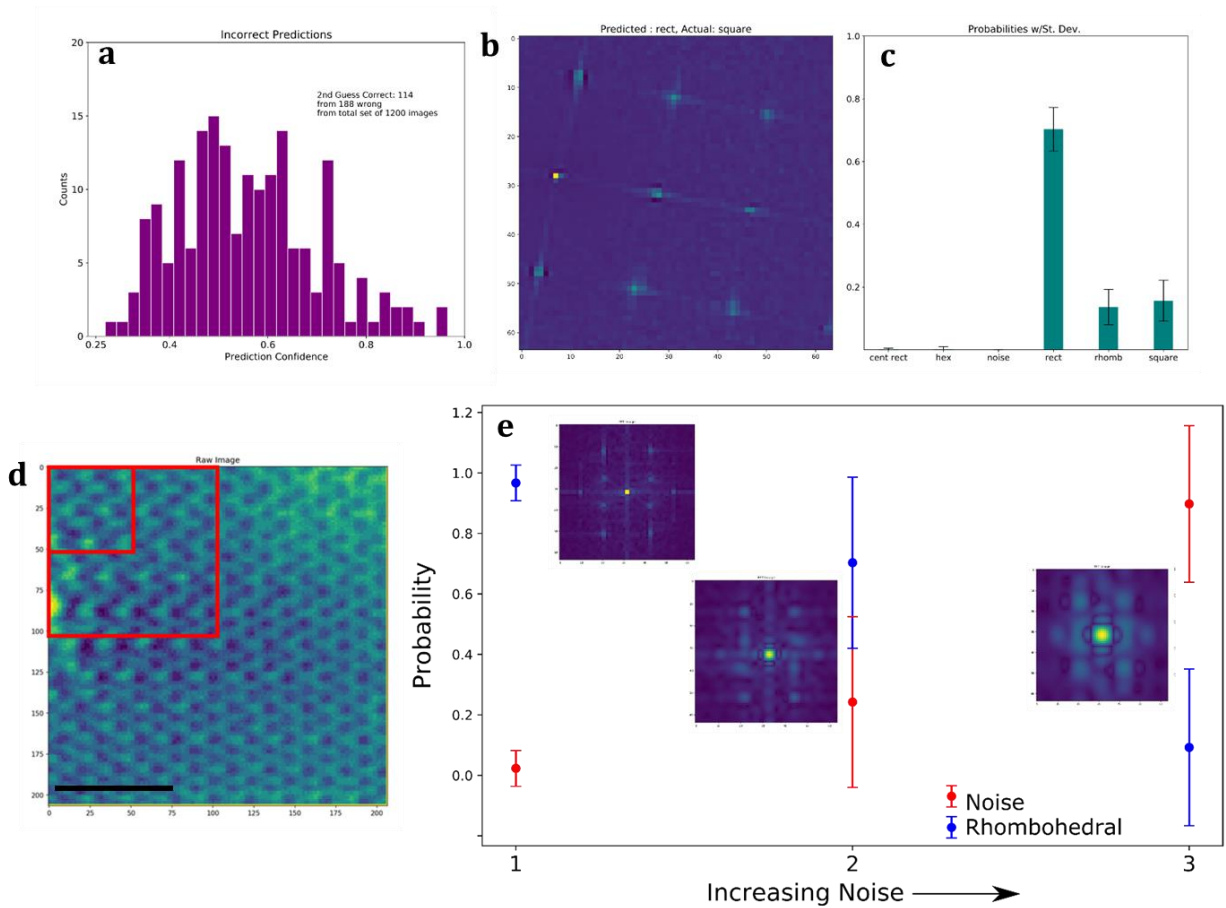

**Figure 3: DCNN Performance and Noise Limits.** (a) Prediction confidence for all incorrect predictions made from a set of (simulated) 1200 images. In total, 188 images were classified incorrectly, and the second-most probable class was correct on 118 of those occasions. (b,c) Single (simulated) test image (b) with the network output for this image in (c). The bars in (c) correspond to one standard deviation, computed via 5000 passes of the image through the network with dropout active. (d) Raw STM image of graphene (scale bar, 1nm). A 2D FFT was performed with the whole image, and in smaller windows (shown in red). The prediction probability for the rhombohedral and 'noise' class for the three windows is shown in (e), with one the error bars marking one standard deviation in the prediction confidence. As the window size becomes smaller, the prediction probability drops.

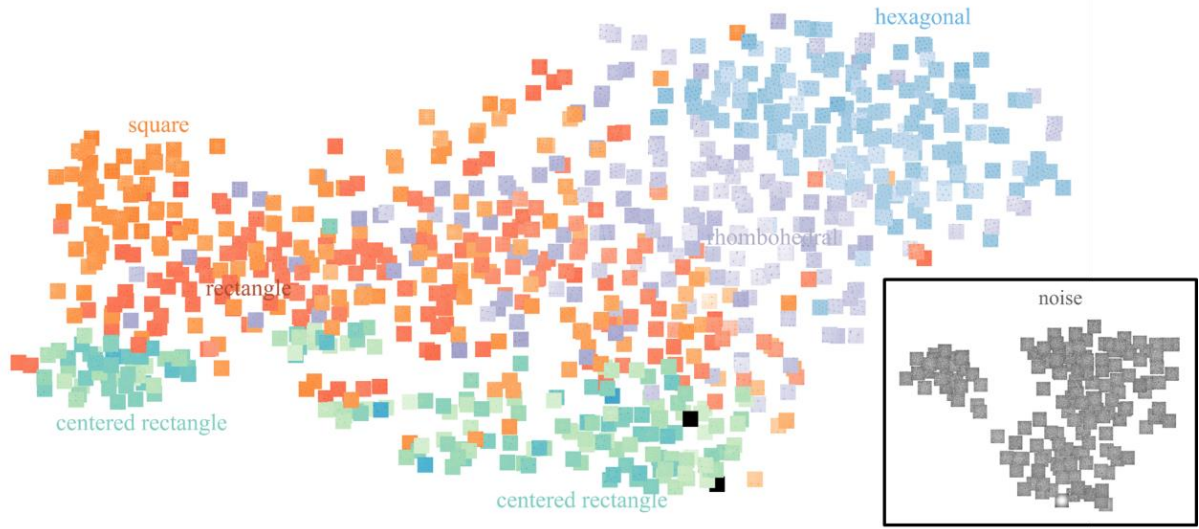

**Figure 4: t-SNE reduced representation of the classes.** The embeddings for 1000 training images are visualized. Essentially, this plot shows the separation between individual images within the classifier. The noise class is in reality much further apart than drawn (thus it has been boxed).

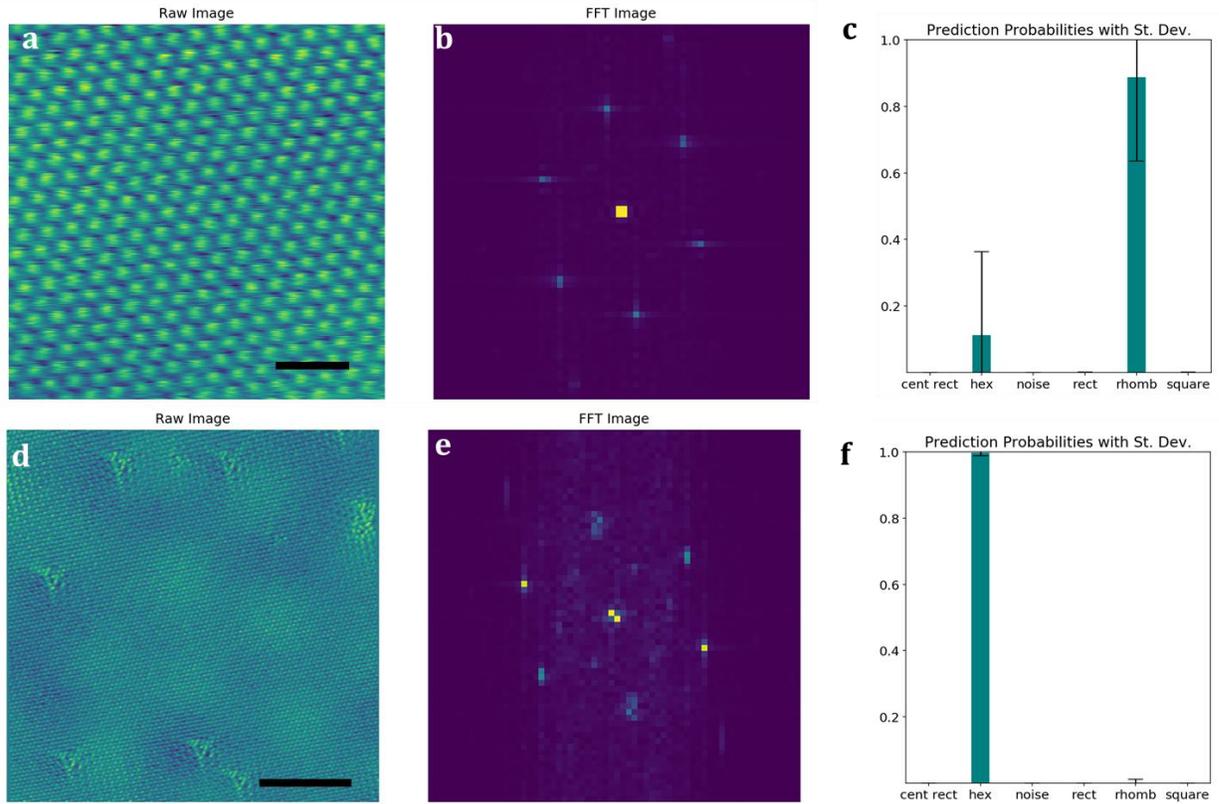

**Figure 5: Application to edge cases of real STM images of graphene.** (**a**) Graphene STM image (scale bar, 1nm) with some slight microscope drift during the scan. (**b**) Associated 2D FFT from image in (a). (**c**) DCNN output for this image. Because of the slight drift, the DCNN suggests that the rhombohedral symmetry is more likely than the hexagonal symmetry, although the variance is large. (**d**) Raw STM Image with both defects and Moiré patterns (scale bar, 5nm). (**e**) 2D FFT of image in (d). (**f**) DCNN output for this image. The hexagonal class is most likely according to the network. Even in the presence of additional reflections from the Moiré pattern, the DCNN is still able to make the correct determination of the symmetry.

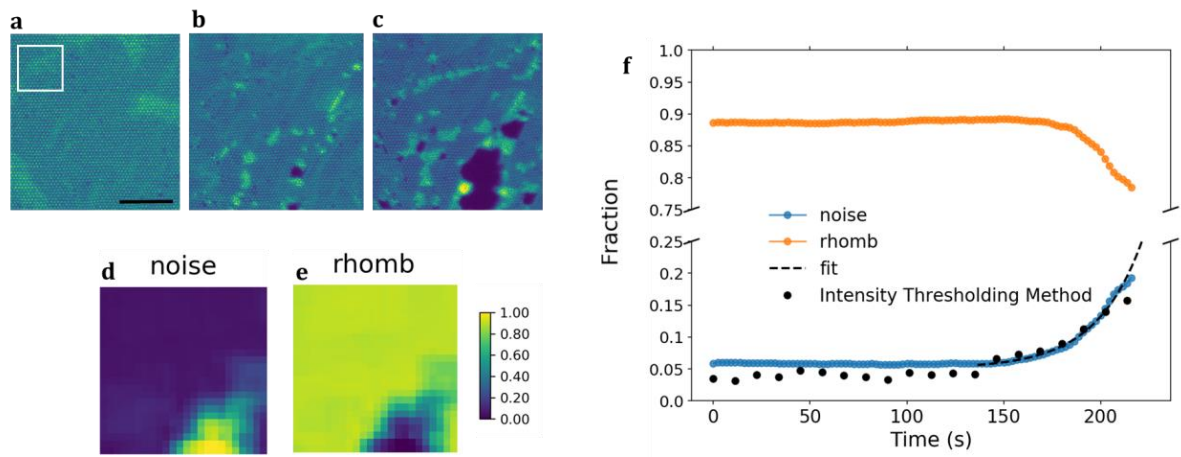

**Figure 6: Analysis of defect evolution in Mo-doped WS₂.** (a-c) STEM images of WS$_2$ as a function of time, at **(a)** 0s, **(b)** 157.5s and **(c)** 202.5s. Scale bar, 5nm. **(d,e)** Probability of classification of the 'noise' (d) and 'rhombohedral' (e) classes for the image in (c). Note that these images are generated via sliding a window (size depicted in (a)) across the image, and inputting each of the windowed images to the DCNN to extract class probabilities. **(f)** Average phase fraction calculated in each frame by the DCNN method (colored lines), and manual intensity thresholding (black circles). An exponential fit (discussed in text) to the data is shown as a black dashed line.

*Supplementary Information*

# Mapping mesoscopic phase evolution during e-beam induced transformations via deep learning of atomically resolved images

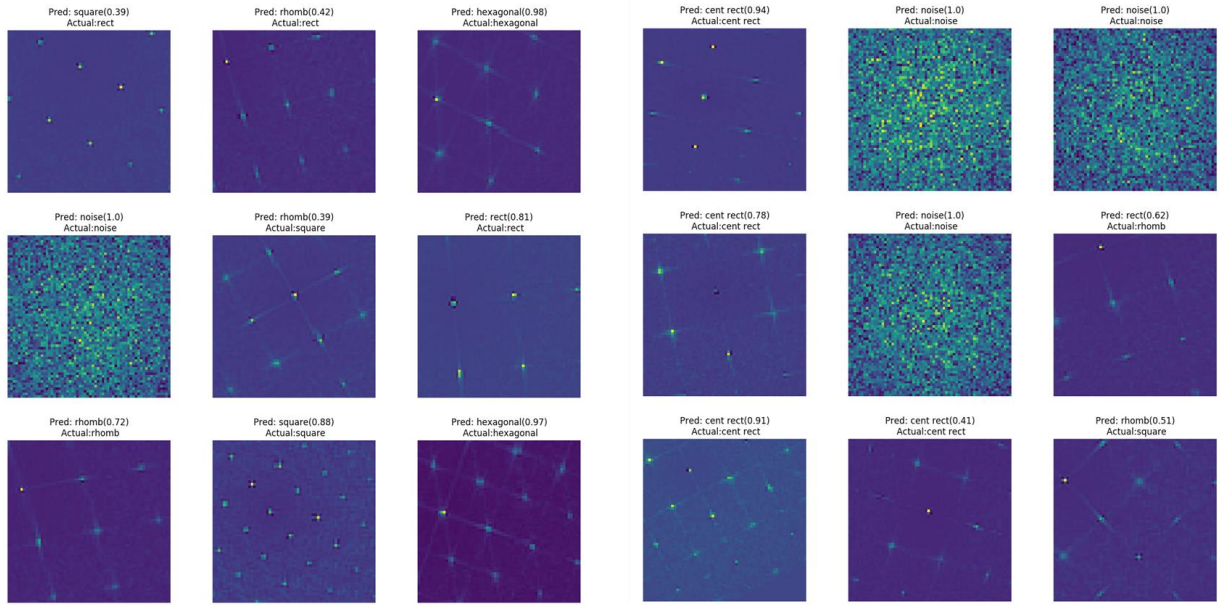

**Supplementary Figure 1:** More examples of the DCNN output on the validation set.

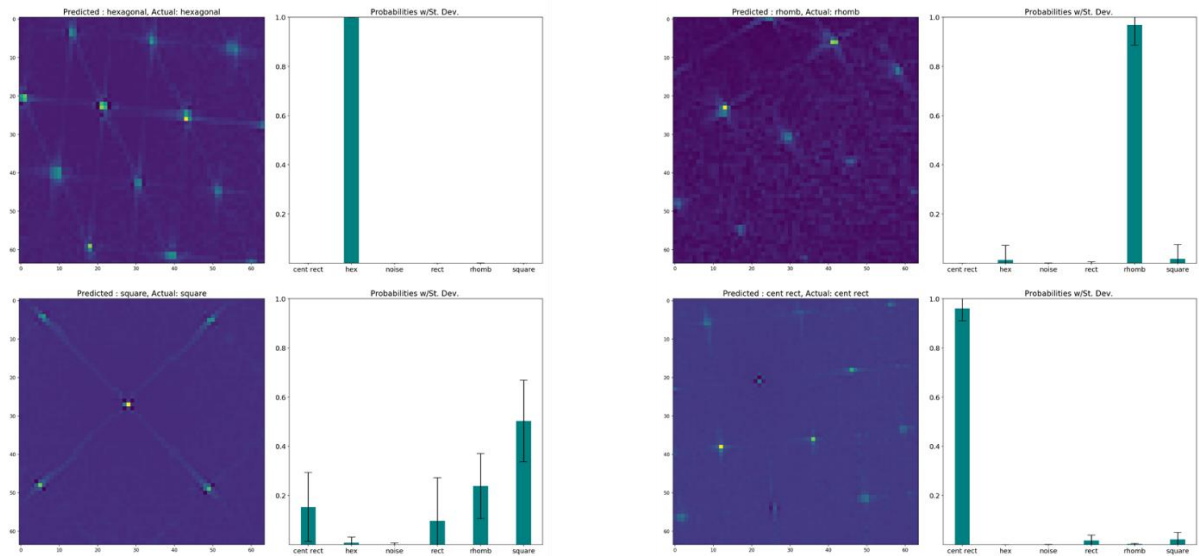

**Supplementary Figure 2:** More examples of the DCNN output on individual images, with the probability over the classifications and one standard deviation shown in the bar charts next to each test image.

**Supplementary Video 1**

The file "predictions_movie.avi" contains a movie of the prediction probabilities for all the classes, for all frames in the movie analyzed in the manuscript within Figure 6.

**Keras Model and Basic Notebook**

The trained Keras model is included with this submission. A basic Jupyter Notebook is included that shows how the model can be used to predict classes from real-space images.